\documentclass[aip,jcp]{revtex4-1}

\usepackage{amsmath}
\usepackage{graphicx}
\usepackage{color}

\newcommand{\qvec}{{\bf q}}
\newcommand{\pvec}{{\bf p}}
\newcommand{\Qvec}{{\bf Q}}
\newcommand{\Pvec}{{\bf P}}

\newcommand{\vvec}{{\bf v}}
\newcommand{\cvec}{{\bf c}}
\newcommand{\fvec}{{\bf f}}
\newcommand{\avec}{{\bf a}}
\newcommand{\zvec}{{\bf z}}

\newcommand{\goo}{g_{\mbox{\tiny OO}}}
\newcommand{\goh}{g_{\mbox{\tiny OH}}}
\newcommand{\ghh}{g_{\mbox{\tiny HH}}}

\begin{document}

\title{
An efficient time-stepping scheme for {\it ab initio} molecular dynamics simulations
}

\author{Eiji  Tsuchida \\ 
  Research Center for Computational Design of Advanced Functional Materials, \\
  National Institute of Advanced Industrial Science and Technology (AIST), \\
  Tsukuba Central 2, Umezono 1-1-1, Tsukuba 305-8568, Japan }

\begin{abstract}
In {\it ab initio} molecular dynamics simulations of real-world problems, 
the simple Verlet method is still widely used for 
integrating the equations of motion, while 
more efficient algorithms are routinely used in classical molecular dynamics. 
We show that if the Verlet method is used in conjunction with pre- and postprocessing, 
the accuracy of the time integration is significantly improved 
with only a small computational overhead. 
The validity of the processed Verlet method is demonstrated 
in several examples including {\it ab initio} molecular dynamics simulations of liquid water. 
The structural properties obtained from the processed Verlet method are found to be sufficiently accurate 
even for large time steps close to the stability limit. 
This approach results in a 2$\times$ performance gain 
over the standard Verlet method for a given accuracy. 
\end{abstract}

\maketitle

\newpage

\section{Introduction}
\label{INTROSEC}

Classical and {\it ab initio} molecular dynamics (AIMD) simulations are 
among the most common methods for theoretical studies of 
complex systems at the atomistic level. 
In the former approach, the interatomic forces are usually given explicitly 
by a sum of pair interactions in analytic form, 
while the latter approach requires quantum-mechanical calculations 
to evaluate the forces \cite{DFTREV1,DFTREV2}. 
AIMD is gaining popularity in the last decades, 
because the accuracy of AIMD is generally higher than the classical approach. 
On the other hand, AIMD is orders of magnitude more expensive, 
which poses a serious obstacle to its wide use in industry. 

In both approaches, the equations of motion for the atoms are 
typically integrated numerically using the symplectic integrators. 
This is because these integrators possess the volume-preserving property in phase space, 
thus leading to the long-term stability of the simulations \cite{TEXT1,TEXT2,TEXT3}. 
In classical molecular dynamics simulations, 
symplectic integrators are often implemented in the form of 
a multiple time step algorithm \cite{STATXT}. 
This algorithm allows us to use large time steps for computationally expensive long-range interactions, 
while inexpensive short-range ones are integrated with small time steps. 
In contrast, single time step symplectic integrators, such as the Verlet method \cite{STATXT,MDTXT1,MDTXT2}, 
are most commonly used in AIMD simulations, 
because the interatomic forces from {\it ab initio} calculations 
cannot be divided into short- and long-range components exactly. 

Theoretical analysis reveals that 
the time step $h$ must satisfy $ h < T / \pi $ in the Verlet method \cite{CLREV1}, 
where $T$ is the period of the fastest motion in the system. 
In many of the classical and {\it ab initio} studies on real-world problems, 
however, the time step satisfies $ h  < T / 15$, 
which is significantly smaller than $T / \pi$. 
This is mainly because the use of a too large time step leads to artifacts 
such as large fluctuations in the total energy and violation of the equipartition theorem \cite{EQP}. 
Nevertheless, it is empirically known that 
molecular dynamics simulations using large time steps often yield surprisingly 
accurate results \cite{LT1,LT2,LT3}. 
Researchers are making constant efforts to understand and exploit 
the robustness of various time integrators 
at large time steps \cite{LT4,LT5,LT6,LT7,LT8,LT9,LT10,FREE1,FREE2,JPS15}. 

In this paper, we investigate a simple extension of symplectic integrators 
called the processing technique. 
This algorithm was originally developed by mathematicians several decades ago \cite{SC1,SC2,SC3,SC4}, 
but has received little attention to date in spite of its solid theoretical background. 
As will be shown below, the processing technique allows us to increase the size of time steps 
by a factor of two at no extra cost, while keeping the accuracy, and thus 
significantly extends the applicability of AIMD simulations \cite{FNOTE0}. 

The rest of the paper is organized as follows. 
In Sec.\ref{THEOSEC}, we present the basic theory of the processed integrator, 
and estimate the computational costs associated with processing. 
Numerical examples demonstrate the effectiveness of this algorithm 
for realistic simulations in Sec.\ref{TESTSEC}. 
In Sec.\ref{DISSEC}, we discuss several possible extensions of the algorithm, 
and present our conclusions.

\section{Theory}
\label{THEOSEC}
\subsection{Verlet method}

The classical Hamiltonian for a system of $N$ atoms is given by 
\begin{equation}
\label{OHAM}
H (\qvec,\pvec) = K (\pvec) + U(\qvec) = \frac12 \, \pvec^T M^{-1} \pvec + U(\qvec), 
\end{equation}
where $\qvec$ and $\pvec$ are vectors of dimension $3N$, 
representing atomic positions and momenta, 
$M$ is the mass matrix, and $U(\qvec)$ is the potential energy. 
Then, the time evolution of any function $A(\qvec, \pvec)$ is given by
\begin{equation}
\frac{dA}{dt} = i L_{\rm H} A = \left\{ A, H \right\}, 
\end{equation}
where the Liouville operator $L_{\rm H}$ is defined by \cite{STATXT} 
\begin{equation}
i L_{\rm H} = \left\{ *, H \right\} =
\sum_i \left( \frac{\partial H}{\partial p_i} \frac{\partial}{\partial q_i}
 - \frac{\partial H}{\partial q_i} \frac{\partial}{\partial p_i} \right). 
\end{equation}
In particular, $\qvec (t)$ and  $\pvec (t)$ satisfy the equations of motion, 
\begin{eqnarray}
\frac{d \qvec}{dt} & = & i L_{\rm H} \qvec = \frac{\partial H}{\partial \pvec}, \\
\frac{d \pvec}{dt} & = & i L_{\rm H} \pvec = -\frac{\partial H}{\partial \qvec}. 
\end{eqnarray}
A formal solution of these equations is given by \cite{STATXT} 
\begin{equation}
\label{EXSOL}
\left(
\begin{array}{c}
\qvec (t)\\
\pvec (t)\\
\end{array}
\right)
= \exp{( i t L_{\rm H})}
\left(
\begin{array}{c}
\qvec (0)\\
\pvec (0) \\
\end{array}
\right) .
\end{equation}
It is easy to prove that the total energy $H(\qvec,\pvec)$ is conserved 
along the trajectory $(\qvec (t),\pvec (t))$, i.e.,
\begin{equation}
\label{EQDHDT}
\frac{d}{dt} H(\qvec,\pvec) = \left\{ H,H \right\} = 0. 
\end{equation}
In general, Eq.(\ref{EXSOL}) cannot be calculated analytically, and thus must be evaluated numerically. 
To this end, $L_{\rm H}$ is first divided into two parts; the kinetic term 
\begin{equation}
i L_1 = \left\{ *,K \right\} = \pvec^T M^{-1} \frac{\partial}{\partial \qvec},  
\end{equation}
and the potential term 
\begin{equation}
i L_2 = \left\{ *,U \right\} = \fvec^T \frac{\partial}{\partial \pvec},  
\end{equation}
where the force is defined by $\fvec = - \partial U / \partial \qvec$. 
Then, using the Suzuki-Trotter decomposition, we obtain 
\begin{equation}
\label{SPLIT}
\exp \left( i h L_{\rm H} \right) = \exp \left( i h (L_1 + L_2) \right) = W_h + O (h^3), 
\end{equation}
where $h$ is the time step, and $W_h$ is given by 
\begin{equation}
\label{L2L1L2}
W_h = \exp \left( \frac{i h L_2}{2} \right) \exp \left( i h L_1 \right) 
\exp \left( \frac{i h L_2}{2} \right). 
\end{equation}
If the $O(h^3)$ terms are neglected, the time evolution of $(\qvec, \pvec)$ can be written as 
\begin{equation}
\label{TIMEEVOL}
\left(
\begin{array}{c}
\qvec_{n+1}\\
\pvec_{n+1}\\
\end{array}
\right)
= W_h 
\left(
\begin{array}{c}
\qvec_{n}\\
\pvec_{n} \\
\end{array}
\right),
\end{equation}
where the subscript denotes the time-step number, i.e., 
$ (\qvec_n, \pvec_n) = (\qvec (nh), \pvec (nh))$. 
The right-hand side of Eq.(\ref{TIMEEVOL}) can be calculated explicitly \cite{STATXT}, 
which leads to the (velocity) Verlet method: 
\begin{eqnarray}
\pvec_{n+\frac12} & = & \pvec_n + \frac{h}{2} \, \fvec_n \\
\qvec_{n+1}       & = & \qvec_n + h M^{-1} \pvec_{n+\frac12}\\
\pvec_{n+1}       & = & \pvec_{n+\frac12} + \frac{h}{2} \, \fvec_{n+1}.
\end{eqnarray}
This integrator is both symplectic and time-reversible. 
These properties are crucial for long-term stability of the simulations \cite{STATXT}. 
Moreover, only one force evaluation is required per step in the Verlet method. 

It is also possible to use higher-order integration schemes \cite{FORU,YSHD,GNS,ODJBCN}, 
which require multiple force evaluations per step. 
These integrators can significantly improve the accuracy for small time steps. 
However, the Verlet method is more robust at large time steps \cite{GNS,ODJBCN} 
which are commonly used in AIMD to minimize the computational cost. 
Therefore, the Verlet method is still the method of choice for AIMD. 
In the following, we focus on the Verlet method, and assume the use of the microcanonical ensemble 
unless otherwise noted. 
Extensions to more general cases will be discussed in Sec.\ref{DISSEC}.

\subsection{Shadow Hamiltonian}
When the Verlet method is used to follow the time evolution of the system, 
the Hamiltonian is no longer a constant of motion, because Eq.(\ref{EQDHDT}) holds only approximately. 
However, any symplectic integrator, including the Verlet method, 
is known to possess a conserved quantity called the shadow (or modified) 
Hamiltonian $H_{\rm S} (\qvec, \pvec)$ \cite{BOND,SIREV1}. 
While the explicit form of $H_{\rm S}$ is not known in general, 
a series expansion in powers of $h$ is valid under mild assumptions \cite{BOND,SIREV1}: 
\begin{equation}
\label{HSEQ}
H_{\rm S}(\qvec,\pvec) = H (\qvec,\pvec) + h^2 H_{(2)} (\qvec,\pvec) + h^4 H_{(4)} (\qvec,\pvec) + \cdots
\end{equation}
In particular, the lowest-order term $H_{(2)}$ 
corresponding to the Verlet method is given by \cite{BOND,SIREV1}
\begin{equation}
\label{H2EQ}
H_{(2)} (\qvec, \pvec) = \frac{1}{12} \, \pvec^T M^{-1} {\cal H} M^{-1} \pvec
-\frac{1}{24} \, \fvec^T M^{-1} \fvec,  
\end{equation}
where the Hessian matrix ${\cal H}$ is defined by 
${\cal H}_{ij}={\partial^2 U}/{\partial q_i}{\partial q_j}$, and 
the Hessian-vector product (${\cal H} M^{-1} \pvec$) can be calculated 
according to Appendix \ref{HVSEC}. 
Higher-order terms are also available in the literature \cite{LSS2}. 

\begin{figure}
  \begin{center}
  \includegraphics[width=8cm]{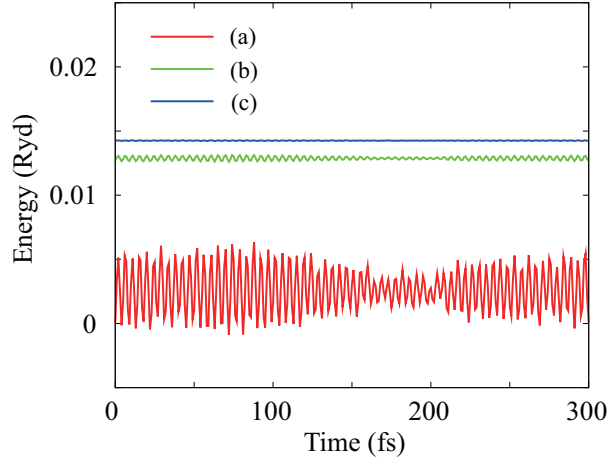}
  \end{center}
  \caption{Time evolution of (a) $H (\qvec_n,\pvec_n)$, (b) $H + h^2 H_{(2)}$, 
    and (c) $H + h^2 H_{(2)} + h^4 H_{(4)}$, 
    for a classical model of water using a time step of 1 fs 
    (see Sec.\ref{CLMDSEC}). 
    $H(\qvec_0,\pvec_0)$ is chosen as the origin, 
    and only up to the second derivatives of $U (\qvec)$ are 
    taken into account when calculating $H_{(4)}$.}
  \label{SHDFIG}
\end{figure}

In Fig.\ref{SHDFIG}, we show the time evolution of 
the original and shadow Hamiltonians for a classical model of water. 
The expansion of the shadow Hamiltonian truncated at $O(h^2)$ shows 
much better conservation than the original Hamiltonian. 
Further improvement is observed if the $O(h^4)$ term is included. 
As is evident from this example, higher-order expansion of the shadow Hamiltonian is 
essentially a constant of motion if the interatomic forces are sufficiently accurate. 
Therefore, the shadow Hamiltonian is often used to detect the errors in forces \cite{BEA1,BEA2,BEA3} 
caused by, e.g., inappropriate truncation of interactions. 
Alternatively, the shadow Hamiltonian can be used to improve the performance of 
the hybrid Monte Carlo method \cite{HMC1,HMC2,HMC3,HMC4}. 
However, as will be shown below, 
we can go a step further and take advantage of 
the shadow Hamiltonian to construct an integrator 
which provides a more accurate trajectory for a given time step.

\subsection{Processed integrator}
\label{PINTSEC}
\begin{figure}
  \begin{center}
  \includegraphics[width=8cm]{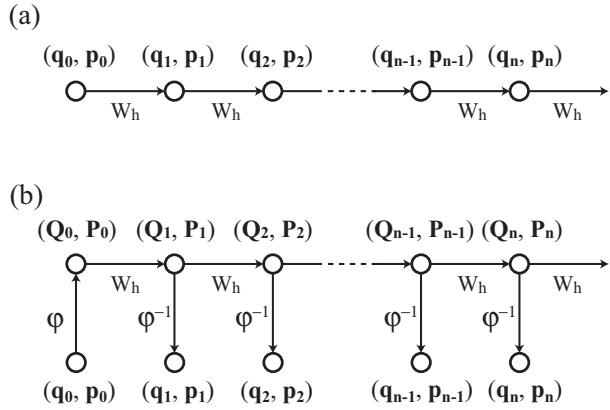}
  \end{center}
  \caption{Flow diagram of the standard (a) and processed (b) symplectic integrators, 
  where $\varphi = \exp{(i h L_{\chi})}$ denotes the preprocessor. }
  \label{FCHART}
\end{figure}

If we invert Eq.(\ref{HSEQ}), we obtain 
\begin{equation}
H (\qvec_n, \pvec_n) = {\rm const.} - h^2 H_{(2)} (\qvec_n, \pvec_n) +O(h^4), 
\end{equation}
which implies that the fluctuations of the total energy are dominated by $H_{(2)}$. 
Here we show how to construct an accurate integrator 
by direct optimization of $H_{\rm (2)}$ through the introduction 
of pre- and postprocessing \cite{SC1,SC2,SC3,SC4}. 
This approach allows us to minimize the effect of 
time-step size, while the computational cost per time step remains the same. 
In this approach, the time evolution of $(\qvec_n,\pvec_n)$ is calculated in three steps: 
\begin{enumerate}
\item Preprocessing is defined by a symplectic transformation of the form 
\begin{equation}
\label{PREPRO1}
\left(
\begin{array}{c}
\Qvec_{n}\\
\Pvec_{n}\\
\end{array}
\right)
= \exp ( i h L_{\chi}) 
\left(
\begin{array}{c}
\qvec_{n}\\
\pvec_{n} \\
\end{array}
\right),
\end{equation}
where 
\begin{equation}
\label{LLDEF}
i L_{\chi} = \left\{ *, H_{\chi} \right\}, 
\end{equation}
and $H_{\chi}$ is the auxiliary Hamiltonian to be described later. 
$\Qvec_n$ and $\Pvec_n$ are intermediate variables with no physical meaning. 
\item Time integration is performed as follows: 
\begin{equation}
\label{SINTEG}
\left(
\begin{array}{c}
\Qvec_{n+1}\\
\Pvec_{n+1}\\
\end{array}
\right)
= W_h 
\left(
\begin{array}{c}
\Qvec_{n}\\
\Pvec_{n} \\
\end{array}
\right),
\end{equation}
where $W_h$ is a symplectic approximation to $\exp (i h L_{\rm H})$, given, e.g., by Eq.(\ref{L2L1L2}). 
\item Postprocessing is the inverse of preprocessing:  
\begin{equation}
\label{POSTPRO1}
\left(
\begin{array}{c}
\qvec_{n+1}\\
\pvec_{n+1}\\
\end{array}
\right)
= \exp (- i h L_{\chi}) 
\left(
\begin{array}{c}
\Qvec_{n+1}\\
\Pvec_{n+1} \\
\end{array}
\right).
\end{equation}
\end{enumerate}
The entire propagator $\Psi$ can be expressed as 
\begin{equation}
\label{PSIEQ}
\Psi = \exp{(-i h L_{\chi})} W_h \exp{(i h L_{\chi})}, 
\end{equation}
which also preserves the symplectic structure. Then, we can easily show that 
\begin{eqnarray}
\left(
\begin{array}{c}
\qvec_{n}\\
\pvec_{n}\\
\end{array}
\right)
& = & \Psi 
\left(
\begin{array}{c}
\qvec_{n-1}\\
\pvec_{n-1}\\
\end{array}
\right)
= \Psi^n 
\left(
\begin{array}{c}
\qvec_{0}\\
\pvec_{0}\\
\end{array}
\right) \nonumber \\
 & = & \exp{(-i h L_{\chi})} W_h^{n} 
\left(
\begin{array}{c}
\Qvec_{0}\\
\Pvec_{0}\\
\end{array}
\right).
\end{eqnarray}
When an initial set of $(\qvec_0, \pvec_0)$ is given, 
we first calculate $(\Qvec_0, \Pvec_0)$ according to Eq.(\ref{PREPRO1}). 
Then, we follow the time evolution of $(\Qvec_0, \Pvec_0)$ 
by repeatedly applying Eq.(\ref{SINTEG}), which yields $(\Qvec_n, \Pvec_n)$ for $n=1,2, \cdots$. 
Only when output is required, we calculate $(\qvec_n, \pvec_n)$ using Eq.(\ref{POSTPRO1}). 
The flow of the algorithm is shown in Fig.\ref{FCHART}. 

At this point, we discuss the choice of $H_{\chi}$ which appears in the definition 
of pre- and postprocessing. Assuming that the Verlet method is used for time integration, 
we adopt the form \cite{LSS1,LSS2} 
\begin{equation}
\label{PPDEF}
H_{\chi} = h \lambda \frac{\partial K}{\partial \pvec} \frac{\partial U}{\partial \qvec}
= h \lambda \pvec^T M^{-1} \frac{\partial U}{\partial \qvec},
\end{equation}
where $\lambda$ is an arbitrary constant \cite{FNOTE1}. 
Then, the lowest-order term of the shadow Hamiltonian corresponding to 
the processed Verlet method of Eq.(\ref{PSIEQ}) is 
given by \cite{SIREV1,LSS1,LSS2} 
\begin{equation}
H_{(2)} (\lambda) = \left(\frac{1}{12} - \lambda\right) \, \pvec^T M^{-1} {\cal H} M^{-1} \pvec
+\left(\lambda - \frac{1}{24}\right) \, \fvec^T M^{-1} \fvec. 
\end{equation}
The value of $\lambda$ should be chosen to minimize the fluctuations of $H_{(2)}(\lambda)$. 
To this end, we assume that $U(\qvec)$ is a quadratic function of $\qvec$, 
\begin{equation}
U (\qvec) = \frac12 \qvec^T U_0 \qvec, 
\end{equation}
where $U_0$ is a constant matrix. 
Then, using $\fvec = -U_0 \qvec$ and ${\cal H} = U_0$, we can prove that 
\begin{equation}
\frac{d}{dt} H_{\rm (2)} (\lambda) = \frac{16 \lambda - 1}{4} \, \pvec^T M^{-1} U_0 M^{-1} U_0 \qvec.  
\end{equation}
Thus, if we choose $\lambda = 1/16$, $ d H_{\rm (2)} / dt = 0$ will hold, 
which is considered the optimal choice in terms of 
total energy conservation \cite{SIREV1,LSS1,LSS2}. 
In the following, the same value of $\lambda$ will be used for 
more general potential functions, including AIMD simulations.

\subsection{Computational cost}
\label{COSTSEC}

By construction, the processed integrator presented in the previous section 
improves the conservation of the total energy for a given time step, or, 
alternatively, allows the use of a larger time step for a given accuracy. 
However, this advantage would be lost if the overhead of processing were significant. 
Here we compare the computational costs of the Verlet method 
with and without the processing. 

When the standard Verlet method is used to integrate the equations of motion, 
we obtain $(\qvec_n, \pvec_n)$ and the corresponding values of 
$H(\qvec_n, \pvec_n) = K (\pvec_n) + U (\qvec_n)$ at the expense of one force evaluation per step. 
At first glance, it may appear a significantly more complicated task 
to calculate the values of $(\qvec_n, \pvec_n)$ and $H(\qvec_n, \pvec_n)$ 
at each step of the processed Verlet method. 
As will be shown below, however, these values can be obtained without 
additional effort if the postprocessing is carefully implemented. 

We first note that the cost of the time-stepping procedure, Eq.(\ref{SINTEG}), is equal to 
the underlying integrator, i.e., one force evaluation ($\fvec (\Qvec_n)$). 
Moreover, the preprocessing, Eq.(\ref{PREPRO1}), needs to be performed 
only once to calculate $(\Qvec_0, \Pvec_0)$ at the beginning of the simulation, 
and is canceled out by the postprocessing in subsequent steps. 
Therefore, the computational cost of preprocessing can be safely ignored. 
The numerical implementation of the preprocessing is presented in Appendix \ref{PREPSEC}. 
We also note that when we start from random initial conditions, 
preprocessing may even be omitted altogether \cite{SIREV1}. 

At variance with preprocessing, the postprocessing is required much more frequently, 
and thus the numerical integration of Eq.(\ref{POSTPRO1}) is prohibitive. 
Instead, we rely on a truncated series expansion in powers of $h$ \cite{LSS1}: 
\begin{eqnarray}
\label{QPOST}
\qvec_n & = & \Qvec_n + h^2 \lambda M^{-1} \fvec (\Qvec_n) + O(h^4), \\
\label{PPOST}
\pvec_n & = & \Pvec_n + h^2 \lambda {\cal H}(\Qvec_n) M^{-1} \Pvec_n + O(h^4). 
\end{eqnarray}
While this procedure preserves the symplectic structure only approximately, 
the errors in $(\qvec_n, \pvec_n)$ do not accumulate with $n$, 
as is evident from Fig.\ref{FCHART}. 
Therefore, the use of Eqs.(\ref{QPOST}) and (\ref{PPOST}) 
does not affect the long-term stability of the integrator. 
Moreover, the error terms of $O(h^4)$ are smaller than those of the Verlet method. 
We also note that the cost of evaluating Eq.(\ref{QPOST}) is negligible, 
since $\fvec (\Qvec_n)$ has already been calculated in the time-stepping procedure. 
Therefore, the values of $\qvec_n$ are available at no extra cost, 
which may be used to calculate various properties such as the radial distribution functions. 

In contrast, a Hessian-vector product is required to calculate $\pvec_n$ using Eq.(\ref{PPOST}). 
The cost of this procedure is comparable to one force evaluation, 
as explained in Appendix \ref{HVSEC}, which offsets the gain 
from the use of a larger time step. 
A simple solution to this problem is to use an alternative expression 
based on the finite-difference approximation \cite{LSS1}: 
\begin{equation}
\label{PPOST2}
\pvec_n = \Pvec_n - \lambda (\Pvec_{n+1} - 2 \Pvec_n  + \Pvec_{n-1}) + O(h^4), 
\end{equation}
which has larger $O(h^4)$ errors, but may be calculated at negligible cost. 
As will be shown in Sec.\ref{TESTSEC}, Eq.(\ref{PPOST2}) is sufficiently accurate 
for {\it a posteriori} analysis of the trajectories obtained from microcanonical simulations. 

In order to obtain the values of $U(\qvec_n)$ at each time step with minimal overhead, 
we propose to use an expansion of the form 
\begin{equation}
\label{UPOST}
U (\qvec_n) = U (\Qvec_n) - h^2 \lambda \, \fvec (\Qvec_n)^T M^{-1} \fvec (\Qvec_n) + O(h^4). 
\end{equation}
This method is similar in spirit to the work of Zhang \cite{ZHANG}. 
In what follows, Eqs.(\ref{PPOST2}) and (\ref{UPOST}) will be referred to 
as the cheap approximations. 

In summary, the processed Verlet method allows us 
to generate a trajectory ($\qvec_n, \pvec_n$, and $H(\qvec_n,\pvec_n)$) 
at the expense of only one force evaluation per step 
if Eqs.(\ref{QPOST}), (\ref{PPOST2}), and (\ref{UPOST}) are used. 
The accuracy of this method is compared with that of the original method in Sec.\ref{TESTSEC}.

\section{Numerical experiments}
\label{TESTSEC}

\subsection{Harmonic oscillator}
\label{HARMSEC}
\begin{figure}
  \begin{center}
  \includegraphics[width=8cm]{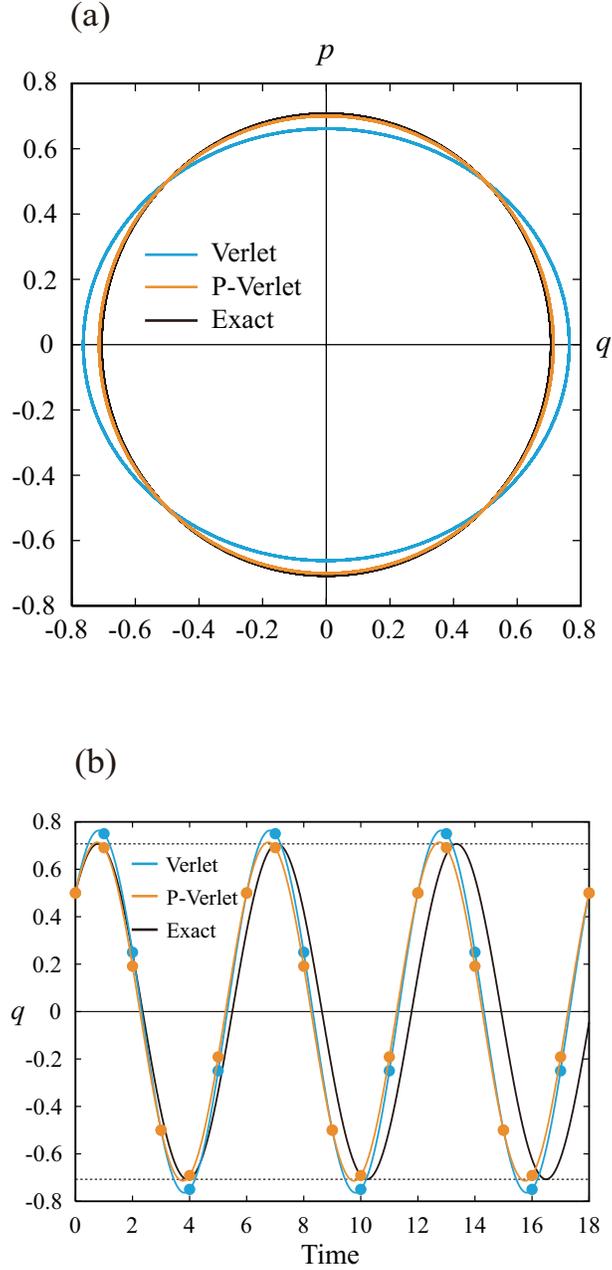}
  \end{center}
  \caption{Numerical results for a one-dimensional harmonic oscillator. 
We assume $m=\omega=h=1$  and $(q_0$, $p_0) = (0.5, 0.5)$. 
(a) Trajectories in phase space for the standard Verlet method 
($\lambda=0$) and the processed Verlet method (P-Verlet, $\lambda=1/16$). 
(b) Time evolution of $q$. Dashed lines denote the theoretical minimum/maximum values 
($= \pm \sqrt{q_0^2 + p_0^2}$).}
  \label{HOFIG}
\end{figure}

We first explore the basic properties of the processed Verlet method 
using a one-dimensional harmonic oscillator which has been 
studied extensively in the past \cite{HO1,HO2}. The Hamiltonian is given by 
\begin{equation}
\label{1DHARM}
H_{\rm 1D} (q, p) = \frac{p^2}{2m} + \frac{m \omega^2}{2} q^2, 
\end{equation}
where $m$ and $\omega$ denote the mass and frequency, respectively. 
Then, from Eq.(\ref{PPDEF}), we obtain 
\begin{equation}
H_{\chi} (q, p) = \lambda h \omega^2 p q. 
\end{equation}
The corresponding preprocessing can be written as 
\begin{equation}
\left(
\begin{array}{c}
Q_{n}\\
P_{n}\\
\end{array}
\right) = \exp (i h L_{\chi})
\left(
\begin{array}{c}
q_{n}\\
p_{n}\\
\end{array}
\right) = 
\left(
\begin{array}{cc}
 \exp{( \lambda h^2 \omega^2)} & 0 \\
 0 & \exp{(-\lambda h^2 \omega^2)} \\
\end{array}
\right)
\left(
\begin{array}{c}
q_{n}\\
p_{n}\\
\end{array}
\right).
\end{equation}
Similarly, the postprocessing is given by 
\begin{equation}
\left(
\begin{array}{c}
q_{n+1}\\
p_{n+1}\\
\end{array}
\right) = 
\left(
\begin{array}{cc}
 \exp{(-\lambda h^2 \omega^2)} & 0 \\
 0 & \exp{( \lambda h^2 \omega^2)} \\
\end{array}
\right)
\left(
\begin{array}{c}
Q_{n+1}\\
P_{n+1}\\
\end{array}
\right). 
\end{equation}
Moreover, the Verlet integrator is given by 
\begin{equation}
\left(
\begin{array}{c}
Q_{n+1}\\
P_{n+1}\\
\end{array}
\right) = W_h 
\left(
\begin{array}{c}
Q_{n}\\
P_{n}\\
\end{array}
\right),
\end{equation}
with
\begin{equation}
W_h = 
\left(
\begin{array}{cc}
1-\frac{h^2 \omega^2}{2} & \frac{h}{m} \\
-hm\omega^2 \left( 1-\frac{h^2 \omega^2}{4} \right) & 1-\frac{h^2 \omega^2}{2}
\end{array}
\right). 
\end{equation}
Now the entire propagator $\Psi_{\rm 1D}$ can be written as 
\begin{equation}
\label{TOTTRANS}
\left(
\begin{array}{c}
q_{n+1}\\
p_{n+1}\\
\end{array}
\right) = \Psi_{\rm 1D}
\left(
\begin{array}{c}
q_{n}\\
p_{n}\\
\end{array}
\right), 
\end{equation}
with 
\begin{eqnarray}
\Psi_{\rm 1D} & = & \exp (- i h L_{\chi}) W_h \exp (i h L_{\chi})  \nonumber \\
 & = & 
\left(
\begin{array}{cc}
1-\frac{h^2 \omega^2}{2} & \frac{h}{m} \exp{(-2 \lambda h^2 \omega^2)} \\
-hm\omega^2 \left( 1-\frac{h^2 \omega^2}{4} \right) \exp{( 2 \lambda h^2 \omega^2)} & 1-\frac{h^2 \omega^2}{2}
\end{array}
\right),
\end{eqnarray}
which also preserves the symplectic structure \cite{STATXT}. 
The exact shadow Hamiltonian corresponding to $\Psi_{\rm 1D}$ is also available: 
\begin{equation}
\label{HS1D}
H_{\rm S} (q_n, p_n) = \frac{p_n^2}{2m} + \beta \frac{m \omega^2}{2} q_n^2 = {\rm const.},
\end{equation}
with 
\begin{equation}
\beta = \left( 1 - \frac{h^2 \omega^2}{4} \right) \exp{\left( 4 \lambda h^2 \omega^2 \right)}
= 1 + 4 \left( \lambda - \frac{1}{16} \right)  h^2 \omega^2 + O (h^4).
\end{equation}
If we choose $\lambda = 1/16$, the $O(h^2)$ terms of $\beta$ vanish, 
regardless of the value of $\omega$, 
thus minimizing the violation of energy equipartition 
arising from the use of a finite time step. 
On the other hand, if $\lambda = 0$ is used, Eq.(\ref{HS1D}) reduces to 
the well-known formula for the standard Verlet method \cite{FREE2,HO1,HO2}. 
In Fig.\ref{HOFIG}(a), we compare the trajectories in phase space 
with and without the processing. The trajectory of the processed Verlet method is 
nearly indistinguishable from the exact solution. 

We now turn to the dynamical behavior of $(q_n, p_n)$ generated by the processed Verlet method. 
The eigenvalues of $\Psi_{\rm 1D}$ can be written as $\exp{(\pm i h \tilde \omega)}$, 
where the modified frequency $\tilde \omega$ is defined by 
\begin{equation}
\label{TRANSOMG}
\tilde \omega = \frac2h \arcsin \left( \frac{h \omega}{2} \right). 
\end{equation}
This result is the same as that for the standard Verlet integrator \cite{HO1,HO2}. 
In particular, $\tilde \omega$ does not depend on $\lambda$, which 
implies that the use of processing does not improve the dynamics \cite{SIREV1}. 
This is also evident from Fig.\ref{HOFIG}(b), 
where the errors in the amplitude are significantly reduced, 
while the phase errors remain uncorrected. 
We note in passing that Eq.(\ref{TRANSOMG}) imposes a limit on the maximum size of $h$, 
as mentioned in Sec.\ref{INTROSEC}. 
Since $\tilde \omega$ is a real number, $h$ must satisfy $ h \omega < 2$, 
or, equivalently, 
\begin{equation}
h < \frac{T}{\pi},
\end{equation}
where the period $T$ is defined by $T = 2 \pi / \omega$.

\subsection{Liquid water: a classical model}
\label{CLMDSEC}
To demonstrate the effectiveness of the processing technique 
in more realistic problems, 
we have performed classical molecular dynamics simulations of liquid water under various conditions. 
Liquid water was modeled by 125 water molecules in a cubic supercell of length 15.67 \AA, 
which corresponds to the density at 353 K. 
The molecular interaction was described by the SPC/Fw force field \cite{SPCFW}, 
which is based on a flexible point-charge water model. 
All interactions were truncated beyond a cutoff distance of 14 Bohr 
using a quintic switching function \cite{SWF1,SWF2}. 
The equations of motion were integrated 
with the standard and processed Verlet methods 
using time steps of 0.3, 0.6, ..., 2.1 fs. 
The simulation was also marginally stable at $h=$ 2.4 fs, 
which corresponds to one-quarter of the period of the O-H stretching motion (9-10 fs). 
However, this case was excluded from the analysis, because 
the total energy exhibited a significant drift as well as large fluctuations. 
We also note that the use of processing has no effect on the maximum stability limit, 
which is dominated by the underlying integrator, i.e., the Verlet method. 
All simulations were started from the same initial conditions $(\qvec_0, \pvec_0)$, 
which were obtained after equilibration, 
and lasted for 1.2 ns in the microcanonical ensemble. 
Preprocessing was performed numerically exactly, 
while the postprocessing was approximated as discussed in Sec.\ref{COSTSEC}. 

\begin{figure}
  \begin{center}
  \includegraphics[width=7cm]{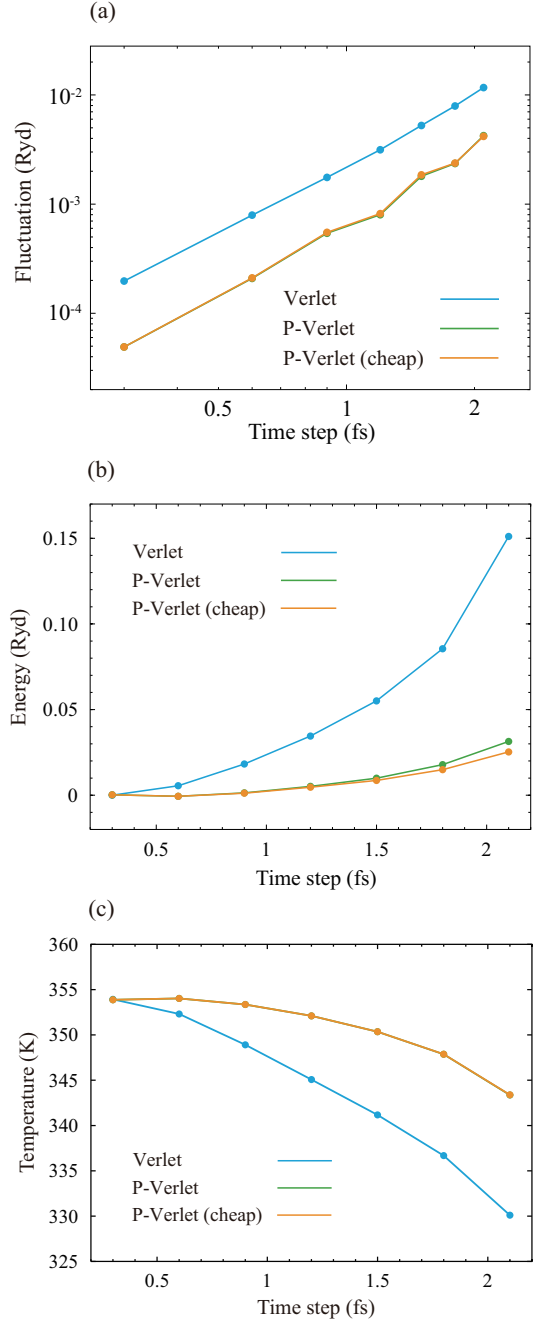}
  \end{center}
  \caption{(a) Fluctuations of the total energy, 
    (b) average values of the potential energy, and 
    (c) average values of the temperature, for a classical model of water. 
    Orange lines denote the results from the processed Verlet method 
    using the cheap approximations for the postprocessing. 
    Green and orange lines are indistinguishable in (a) and (c), indicating 
    the accuracy of the cheap approximations. 
    The potential energy from the standard Verlet method at $h = 0.3$ fs 
    is chosen as the origin in (b). 
  }
  \label{NACLWFIG}
\end{figure}

We first investigate the accuracy of numerical integration 
with and without the processing. 
In Fig.\ref{NACLWFIG}, we show the fluctuations of the total energy, 
average potential energies, and average temperatures, each as a function of time step. 
The fluctuations were calculated as the standard deviation from a linear fit to the total energy. 
The drift was negligibly small in all runs. 
Figure \ref{NACLWFIG} suggests that the numerical accuracy of the Verlet method 
using a time step of $h$ is comparable to that of the processed Verlet method using $2h$. 
Moreover, the effect of cheap approximation is found to be small 
for both atomic momentum (Eq.(\ref{PPOST2})) and potential energy (Eq.(\ref{UPOST})). 
Therefore, the processed Verlet method is more efficient than the original method by about a factor of two, 
in agreement with previous studies \cite{LSS1}. 

\begin{figure}
  \begin{center}
  \includegraphics[width=8cm]{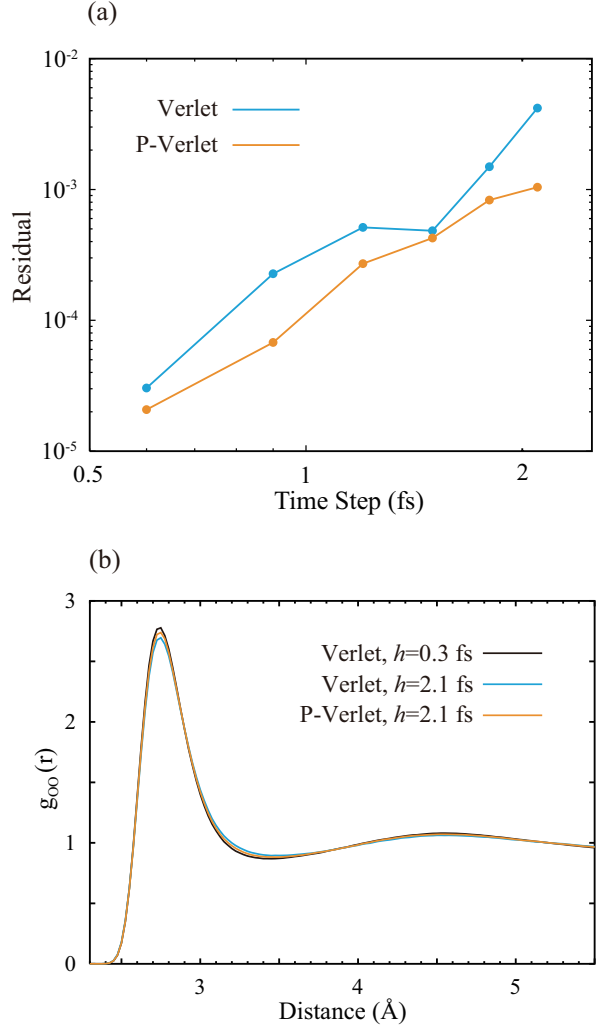}
  \end{center}
  \caption{(a) Errors in the radial distribution functions, Eq.(\ref{RSDLEQ}), 
    for a classical model of water. 
    (b) $\goo^{\rm ref} (r)$ and $\goo (r)$ for $h=2.1$ fs with and without the processing. 
  }
  \label{RDFFIG}
\end{figure}

Now we compare the structural properties obtained from each run. 
In Table \ref{BONDTBL1}, we show the average lengths of covalent O-H bonds, 
together with their fluctuations. 
The average values are found to be insensitive to $h$, 
while the fluctuations show an increase of 10 \% at large time steps for the Verlet method. 
In contrast, all results from the processed Verlet method agree within statistical errors. 
The intermolecular structure is described by $\goo (r)$, $\goh (r)$, and $\ghh (r)$, 
representing oxygen-oxygen, oxygen-hydrogen, and hydrogen-hydrogen 
radial distribution functions, respectively. 
The error caused by the finite size of the time step is defined by 
\begin{equation}
\label{RSDLEQ}
R (h) = \int_0^{r_{\rm max}} dr \left(
\left|\goo^h (r) - \goo^{\rm ref} (r)\right|^2 +
\left|\goh^h (r) - \goh^{\rm ref} (r)\right|^2 +
\left|\ghh^h (r) - \ghh^{\rm ref} (r)\right|^2 \right), 
\end{equation}
where we set $r_{\rm max}$ = 7.5 \AA, and the reference values 
($\goo^{\rm ref}$, $\goh^{\rm ref}$, and $\ghh^{\rm ref}$) are the results for $h=0.3$ fs without processing. 
The residual errors shown in Fig.\ref{RDFFIG}(a) suggest that 
although the processed Verlet method is more accurate than the standard Verlet method, 
the improvement is relatively small, being comparable to a time step reduction of only 0.3 fs. 
To make this point more explicit, 
we compare $\goo (r)$ for $h=2.1$ fs with and without the processing in Fig.\ref{RDFFIG}(b). 
This figure indicates that the intermolecular structure of this system 
can be described with reasonable accuracy using large time steps near the stability limit, 
even if no processing is applied. 
This is mainly because the intermolecular structure of liquid water is 
relatively insensitive to the intramolecular vibrations, 
as demonstrated by the success of rigid water models \cite{RIGIDW,FNOTE5}. 

\begin{figure}
  \begin{center}
  \includegraphics[width=8cm]{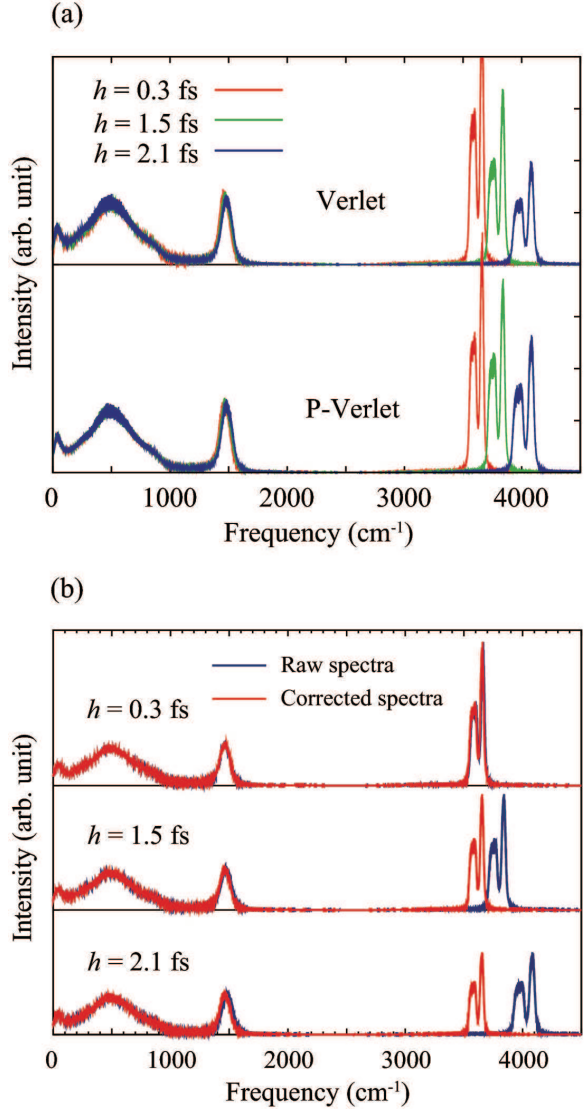}
  \end{center}
  \caption{(a) Power spectra for a classical model of water 
    obtained from the standard and processed Verlet methods. 
    All peaks above 1000 cm$^{-1}$ correspond to intramolecular vibrations. 
    (b) Power spectra from the processed Verlet method
    before and after the correction of Eq.(\ref{TRANSOMG2}).} 
  \label{PSFIG}
\end{figure}

\begin{figure}
  \begin{center}
  \includegraphics[width=8cm]{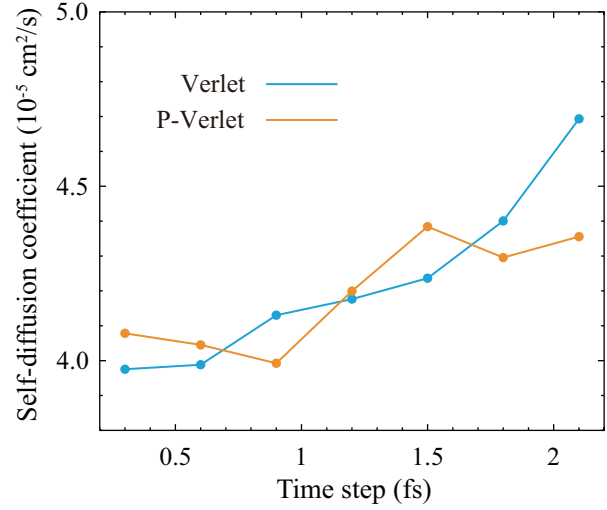}
  \end{center}
  \caption{Self-diffusion coefficients for a classical model of water 
    obtained from the standard and processed Verlet methods.}
  \label{DSELFFIG}
\end{figure}

We now turn to the dynamical properties of this system. 
To this end, we have calculated the power spectra and self-diffusion coefficients, 
as shown in Figs.\ref{PSFIG} and \ref{DSELFFIG}. 
The power spectrum $I(\omega)$ is defined by 
\begin{equation}
I (\omega) = \sum_{i=1}^{N} 
\int_0^{\infty} \left\langle \vvec_i (t) \cdot \vvec_i (0) \right\rangle \cos{(\omega t)} dt, 
\end{equation}
where $\vvec_i (t)$ is the velocity of atom $i$ at time $t$, 
and the angle brackets denote the autocorrelation function. 
The self-diffusion coefficient $D_{\rm self}$ is given by \cite{FNOTE2} 
\begin{equation}
D_{\rm self} = \frac{1}{3N} \sum_{i=1}^{N} 
\int_0^{\infty} \left\langle \vvec_i (t) \cdot \vvec_i (0)\right\rangle dt. 
\end{equation}
As already mentioned in Sec.\ref{HARMSEC}, 
there is no reason to expect that the dynamical properties are improved by 
the use of processing. 
Our results for $I(\omega)$ and $D_{\rm self}$ are consistent with this observation. 
In particular, the high-frequency part of the spectra shown in Fig.\ref{PSFIG}(a)
exhibits a substantial blue shift at large time steps 
regardless of the use of processing, 
while the low-frequency part (below 1000 cm$^{-1}$) remains the same in all cases. 
We have found, however, that the peak positions of the spectra 
in the limit of zero time step can be accurately estimated by a simple correction formula, 
\begin{equation}
\label{TRANSOMG2}
\omega_0 = \frac2h \sin \left( \frac{h \omega}{2} \right),
\end{equation}
which corresponds to the inverse of Eq.(\ref{TRANSOMG}). 
Here $\omega_0$ denotes the frequency at zero time step. 
We compare the power spectra with and without the correction in Fig.\ref{PSFIG}(b), 
where the corrected results show excellent agreement with each other. 
We also note that Eq.(\ref{TRANSOMG2}) is valid 
whether or not the processing is applied. Therefore, this procedure is also useful 
for estimating the correct peak positions in conventional molecular dynamics simulations. 
As is evident from Fig.\ref{DSELFFIG}, 
the self-diffusion coefficient remains nearly constant at small time steps ($h < 1$ fs), 
and grows slowly with $h$ at larger time steps in both methods. 
These results are consistent with the claim that 
the use of processing has no apparent effect on the dynamical properties \cite{SIREV1}.

\subsection{Liquid water: an {\it ab initio} study}
In principle, the processing technique should be equally valid for AIMD simulations. 
However, we are not aware of any previous work in this direction. 
Here we study the effect of processing on the performance of AIMD simulations 
for liquid water. Liquid water at 423 K was modeled 
by 64 molecules in a cubic supercell of length 13.92 \AA \cite{LIQW}. 
Atomic forces were calculated within the density functional theory \cite{HK,KS,PBE}, and 
norm-conserving pseudopotentials were employed \cite{GTH,HGH}. 
Only the $\Gamma$-point was used to sample the Brillouin zone. 
The electronic orbitals were expanded by the 
finite-element basis functions \cite{FEM1,FEM2,FEM3} with 
an average cutoff energy of 58 Ryd, while the resolution was enhanced by about a factor of 2 
near the oxygen atoms by adaptation of the grid \cite{ACC}. 
The electronic states were quenched to the Born-Oppenheimer surface 
at each time step with the limited-memory BFGS method in mixed precision arithmetic \cite{LINO,QNFEM,MIXP}. 
The equations of motion for the atoms were integrated using 
the Verlet method with and without the processing. 

After several preliminary runs, the Verlet method was found to be stable up to $h =$ 1.2 fs, 
while a significant drift in the total energy was observed at $h = 1.5$ fs. 
These values are somewhat smaller than the corresponding values for the classical model, 
even though the highest frequency is nearly the same in both cases. 
This discrepancy is explained by the strong anharmonicity of 
the potential energy surface obtained from {\it ab initio} calculations \cite{PASC}. 
After equilibration, production runs of 30 ps were carried out in the microcanonical ensemble 
using $h =$ 0.3, 0.6, 0.9, and 1.2 fs \cite{FNOTE3}. 
We used the same initial conditions $(\qvec_0, \pvec_0)$ and 
experimental masses for all atoms in these runs. 
For comparison, we also include the results for the processed Verlet method 
using the optimized masses ($m_{\rm H} =$ 4.5, $m_{\rm O} =$ 9) \cite{FEEN,MTMD2} 
and $h =$ 1.0, 1.5, and 2.0 fs. 
When going from the experimental to the optimized masses, 
the highest-frequency peak moves from 3600 cm$^{-1}$ to 2100 cm$^{-1}$. 
Therefore, the value of $h/T$ for $h =$ 1.2 fs in the former case is comparable to 
that for $h =$ 2.0 fs in the latter case. 
The initial conditions for $(\qvec, \pvec)$ were adjusted to give 
the same total energy as for the experimental masses, 
and after re-equilibration, data were collected for 30 ps. 

\begin{figure}
  \begin{center}
  \includegraphics[width=8cm]{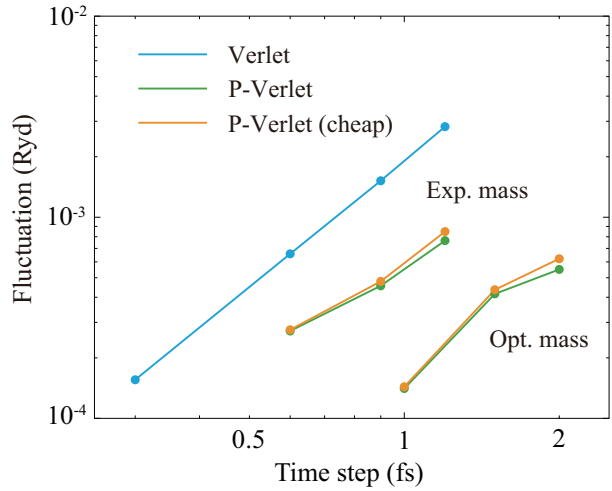}
  \end{center}
  \caption{Fluctuations of the total energy obtained from AIMD simulations of liquid water. \\
  Same notation as in Fig.\ref{NACLWFIG}.}
  \label{AIMD1FIG}
\end{figure}

\begin{figure}
  \begin{center}
  \includegraphics[width=8cm]{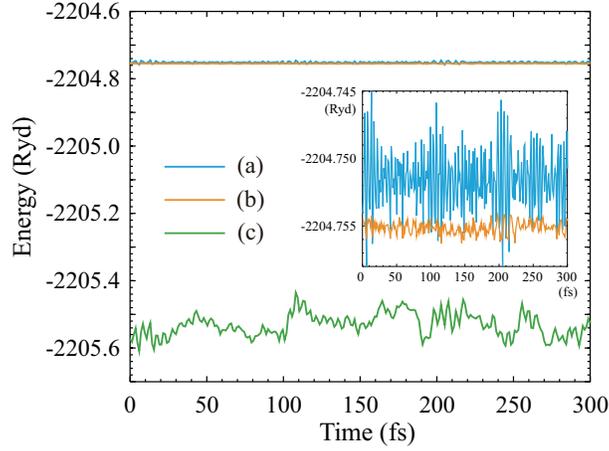}
  \end{center}
  \caption{Time evolution of the total and potential energies in an AIMD simulation 
    of liquid water using $h =$ 1.2 fs: (a) total energy obtained from 
    the standard Verlet method, (b) total energy from the processed Verlet method, 
    and (c) potential energy from the processed Verlet method. 
    The inset is a magnification of the total energies.}
  \label{PV2ENEFIG}
\end{figure}

In Fig.\ref{AIMD1FIG}, we show the fluctuations of the total energy for all runs. 
Although the results show some scatter, the processed Verlet method is 
approximately twice as efficient as the Verlet method, in agreement with the classical case. 
An additional gain is obtained by the mass scaling method \cite{FEEN,MTMD2}. 
In particular, the accuracy of the processed Verlet method using $h =$ 2.0 fs and the optimized masses 
is comparable to that of the standard Verlet method using $h =$ 0.6 fs. 
Moreover, the use of cheap approximations results in only small changes in accuracy. 
Figure \ref{PV2ENEFIG} shows the time evolution of the total and potential energies 
with and without the processing. 

\begin{figure}
  \begin{center}
  \includegraphics[width=8cm]{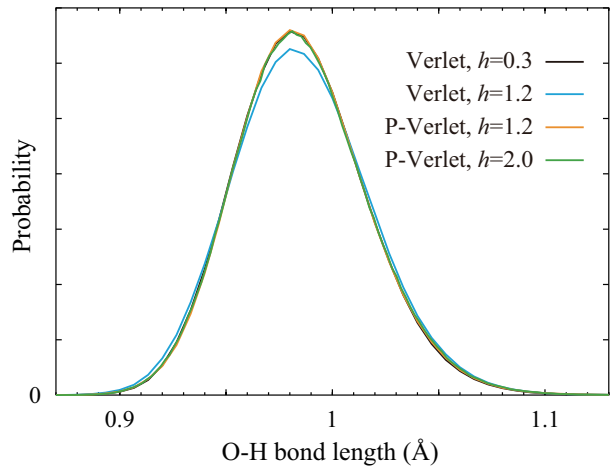}
  \end{center}
  \caption{Probability distributions of O-H bond length 
    obtained from AIMD simulations of liquid water.}
  \label{OHPROBFIG}
\end{figure}

We now compare the average lengths of covalent O-H bonds 
from AIMD simulations in Table \ref{BONDTBL2}. 
Similar to the classical case, 
the average values are insensitive to the size of time steps, while the fluctuations are 
clearly improved by the use of processing. 
The probability distributions of O-H bond length are also shown in Fig.\ref{OHPROBFIG}. 
All curves except that of the standard Verlet method using $h = 1.2$ fs are nearly identical. 

\begin{figure}
  \begin{center}
  \includegraphics[width=8cm]{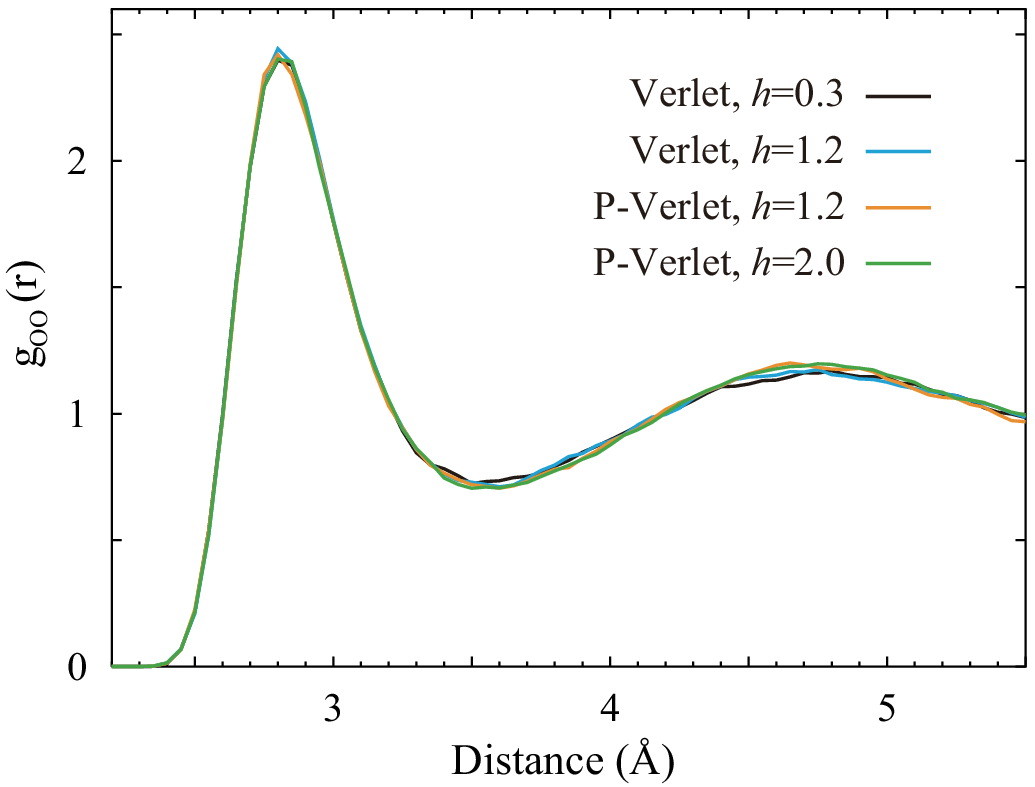}
  \end{center}
  \caption{Oxygen-oxygen radial distribution functions 
    obtained from AIMD simulations of liquid water.}
  \label{DFTGOOFIG}
\end{figure}

Figure \ref{DFTGOOFIG} shows the oxygen-oxygen radial distribution functions in selected cases. 
Unfortunately, the statistical errors due to the limited length of our simulations 
are found to be larger than the systematic errors due to the finite time step, 
particularly at large distances. 
However, all runs give very similar results 
for the first peak at 2.8 \AA. 
This is not surprising considering the small errors 
in the classical case shown in Fig.\ref{RDFFIG}.

\section{Discussion and Conclusions}
\label{DISSEC}

Thus far, we have focused on the implementation of the processed Verlet method 
in the microcanonical ensemble. 
However, the use of a thermostat is often desired in real applications 
to generate the canonical ensemble \cite{HUNEN}. 
Unfortunately, many of the standard thermostats 
require the values of instantaneous temperature \cite{STATXT,MDTXT1,MDTXT2}, and thus 
we need to calculate $\pvec_n$ on-the-fly using Eq.(\ref{PPOST}) at each time step. 
The performance gain from the processed Verlet method is offset 
by the computational overhead of this procedure. 

One possible solution to this problem is to use 
the stochastic thermostat originally developed by Heyes \cite{MDTXT1,HYS}. 
In this approach, the temperature needs to be evaluated and updated less frequently, 
say every 10 steps, and during each interval, 
the equations of motion are integrated in the microcanonical ensemble. 
Therefore, the cost of calculating the instantaneous temperature is reduced to an acceptable level. 
The temperature is updated by scaling all atomic momenta by a common factor $\gamma \approx 1$. 
After the update, we do not need to preprocess the new momenta explicitly 
thanks to Eq.(\ref{PSCALE_EQ}). 
The overhead is further reduced if Eq.(\ref{PPOST2}) is used for calculating $\pvec_n$, 
because the update is often skipped in the Heyes thermostat \cite{HYS}. 
Preliminary results suggest that the canonical ensemble 
can actually be generated with an overhead of 5-10 \%. 

Another possible extension of the present algorithm is the higher-order integrator 
which was first introduced by Rowlands \cite{RLBIB}, 
and later reformulated by L\'opez-Marcos {\it et al.} \cite{LSS1,LSS2}. 
This integrator also consists of three steps, as already mentioned in Sec.\ref{PINTSEC}. 
The time integration step corresponding to Eq.(\ref{SINTEG}) is given by 
\begin{eqnarray}
\label{RWLDS1}
\pvec_{n+\frac12} & = & \pvec_n + \frac{h}{2} \, \left[ \fvec - \alpha h^2 {\cal H} M^{-1} \fvec \right]_n \\
\label{RWLDS2}
\qvec_{n+1}       & = & \qvec_n + h M^{-1} \pvec_{n+\frac12} \\
\label{RWLDS3}
\pvec_{n+1}       & = & \pvec_{n+\frac12} + \frac{h}{2} \, 
\left[ \fvec - \alpha h^2 {\cal H} M^{-1} \fvec \right]_{n+1} 
\end{eqnarray}
with $\alpha = 1/12$ \cite{FNOTE4}. 
The auxiliary Hamiltonian of Eq.(\ref{PPDEF}) should also be extended to include $O(h^3)$ terms \cite{LSS1,LSS2}. 
In order to evaluate the performance of this algorithm, several test calculations 
have been carried out. The results indicate significant improvement of 
accuracy for a given time step compared to the processed Verlet method. 
Moreover, the maximum stability limit is increased by a factor of $\approx$ 1.5. 
On the other hand, the computational cost per time step is 
about twice as expensive as that of the processed Verlet method, 
because we need to calculate a Hessian-vector product at each step. 
Therefore, the increase in time-step size is insufficient to compensate for the overhead, 
and thus the processed Verlet method should be preferred 
in terms of total performance. 
If, however, a highly accurate trajectory is required, 
this approach would be a viable option. 

In summary, the processed Verlet method is about twice as efficient as 
the standard Verlet method if the cheap approximation to the postprocessing is applied. 
We have also shown how to generate the canonical ensemble with only a small overhead. 
This algorithm would be particularly useful for AIMD simulations 
which are not easily compatible with the multiple time step algorithm \cite{STATXT}. 
It is also straightforward to use this algorithm in conjunction with other methods which accelerate 
the electronic structure calculations \cite{AOMM,OZ,ONREV}, 
as well as efficient sampling of the phase space \cite{JCC08}.

\section*{Acknowledgments}
This work has been supported by the Strategic Programs for Innovative Research (SPIRE) 
and a KAKENHI grant (22104001) from the Ministry of Education, Culture, Sports, 
Science \& Technology (MEXT), 
the Japan-US Cooperation Project for Research and Standardization of 
Clean Energy Technologies from the Ministry of Economy, Trade, and Industry (METI), 
and the Computational Materials Science Initiative (CMSI), Japan. 
Part of the calculations were carried out using the computer facilities at Research 
Institute for Information Technology, Kyushu University.

\appendix

\section{Hessian-Vector Product}
\label{HVSEC}
Evaluation of a Hessian-vector product ($\zvec = {\cal H} \cvec$) 
for any given vector $\cvec$ is required, e.g., 
in Eqs.(\ref{H2EQ}), (\ref{PPOST}), and (\ref{RWLDS1}). 
Each element of $\zvec$ can be written as 
\begin{equation}
z_i = \sum_j {\cal H}_{ij} c_j = \sum_j c_j \frac{\partial}{\partial q_j} 
\left(\frac{\partial U}{\partial q_i}\right) 
= - \lim_{\epsilon \rightarrow 0} 
\frac{f_i (\qvec + \epsilon \cvec) - f_i (\qvec - \epsilon \cvec)}{2 \epsilon}, 
\end{equation}
where $\qvec$ denotes the current atomic positions. 
When $U(\qvec)$ is given by a sum of pair interactions, 
$\zvec$ can be calculated analytically, and its computational cost is comparable to 
that of $\fvec (\qvec)$ \cite{LSS1,LSS2}. 
A more complicated procedure is required for AIMD, because the explicit form of $U (\qvec)$ is unknown. 
Let us assume that $U (\qvec)$, $\fvec (\qvec)$, and the corresponding 
ground-state orbitals $\Phi (\qvec)$ have already been calculated. 
Then, $\zvec$ can be evaluated in either of the two ways: 
by the density functional perturbation theory \cite{MBPT1,MBPT2} or by the finite-difference method. 

In the density functional perturbation theory, 
we first obtain the change of $\Phi (\qvec)$ induced by 
an infinitesimal displacement of atoms along $\cvec$, 
denoted by $({\partial \Phi}/{\partial \cvec})$, 
from the iterative solution of the Sternheimer equation \cite{MBPT1,MBPT2,MBPT3}. 
Then, $\zvec$ can be calculated analytically using 
$\Phi (\qvec)$ and $({\partial \Phi}/{\partial \cvec})$ \cite{MBPT3}. 
The computational cost of this procedure is comparable to that of 
calculating $\fvec (\qvec)$ \cite{MBPT3}. 
We note, however, that the numerical implementation of the density functional perturbation theory 
requires significant programming efforts. 

Alternatively, if we calculate the forces at $\qvec \pm \epsilon_0 \cvec$, 
$\zvec$ may be evaluated by the finite-difference method: 
\begin{equation}
\zvec \approx - \frac{\fvec (\qvec + \epsilon_0 \cvec) - \fvec (\qvec - \epsilon_0 \cvec)}{2 \epsilon_0}, 
\end{equation}
where $\epsilon_0$ is a small positive constant. 
At first glance, the second approach may look twice as expensive as the first. 
However, the calculation of $\fvec (\qvec - \epsilon_0 \cvec)$ is much cheaper 
than that of $\fvec (\qvec + \epsilon_0 \cvec)$, 
because an extremely good initial guess for $\Phi (\qvec - \epsilon_0 \cvec)$ is 
available from Eq.(\ref{EXTEQ1}) \cite{PAOMD}. 
Therefore, the total cost of this approach is also comparable to one force evaluation. 
For simplicity, we have adopted the finite-difference approach throughout this study. 

In either case, the directional derivative, 
\begin{equation}
\label{PHIDRV}
\frac{\partial \Phi}{\partial \cvec} = \cvec \cdot \nabla_\qvec \Phi \approx 
\frac{\Phi (\qvec + \epsilon_0 \cvec) - \Phi (\qvec - \epsilon_0 \cvec)}{2 \epsilon_0} 
\end{equation}
is obtained as a by-product, 
which may be utilized to enhance the initial guess, as explained in Appendix \ref{EXTSEC}.

\section{Preprocessing}
\label{PREPSEC}
Here we show how to perform the preprocessing of Eq.(\ref{PREPRO1}) explicitly for $n=0$, 
\begin{equation}
\label{PREPRO_APP}
\left(
\begin{array}{c}
\Qvec_{0}\\
\Pvec_{0}\\
\end{array}
\right)
= \exp{(i h L_{\chi})}
\left(
\begin{array}{c}
\qvec_{0}\\
\pvec_{0}\\
\end{array}
\right). 
\end{equation}
The right-hand side of Eq.(\ref{PREPRO_APP}) can be calculated 
by integrating the equations of motion for $(\qvec (\tau), \pvec (\tau))$, 
\begin{eqnarray}
\label{DQDTEQ}
\frac{d \qvec}{d\tau} & = & \frac{\partial H_{\chi}}{\partial \pvec} 
	= h \lambda M^{-1} \frac{\partial U}{\partial \qvec} \\
\label{DPDTEQ}
\frac{d \pvec}{d\tau} & = & - \frac{\partial H_{\chi}}{\partial \qvec} 
	= - h \lambda {\cal H} M^{-1} \pvec
\end{eqnarray}
from $\tau = 0$ to $\tau = h$, assuming that $(\qvec (0), \pvec (0)) = (\qvec_0, \pvec_0)$ and 
$H_{\chi}$ is given by Eq.(\ref{PPDEF}). 
The Hessian-vector product appearing in Eq.(\ref{DPDTEQ}) can be calculated 
according to Appendix \ref{HVSEC}. 
These equations can be integrated numerically by any standard method \cite{RCP}.  
A fully converged solution 
$(\Qvec_0, \Pvec_0) = (\qvec (h), \pvec (h)) $ is typically obtained 
at the expense of 10-20 force evaluations. 

It is worth noting that when Eq.(\ref{PREPRO_APP}) holds, so does 
\begin{equation}
\label{PSCALE_EQ}
\left(
\begin{array}{c}
\Qvec_{0}\\
\gamma \Pvec_{0}\\
\end{array}
\right)
= \exp{(i h L_{\chi})}
\left(
\begin{array}{c}
\qvec_{0}\\
\gamma \pvec_{0}\\
\end{array}
\right)
\end{equation}
for any constant $\gamma$, because Eqs.(\ref{DQDTEQ}) and (\ref{DPDTEQ}) are separable. 
This property is useful for scaling the atomic momenta 
when the temperature is controlled by a thermostat.

\section{Initial Guess}
\label{EXTSEC}
The computational cost of AIMD simulations is dominated by 
an iterative procedure for calculating the ground-state orbitals 
$\Phi_n = \Phi (\qvec_n)$ for each $n$ \cite{DFTREV1,DFTREV2}. 
Therefore, it is important to start from a good initial guess 
for $\Phi_n$ to minimize the computational effort. 
To this end, we present several possible approximations to $\Phi_{n+1}$, 
denoted by $\Phi^{\rm init}_{n+1}$. 
For simplicity, we limit ourselves to non-metallic systems here, 
and $(\qvec, \pvec)$ should be replaced by $(\Qvec, \Pvec)$ when necessary.  
Assuming that 
\begin{equation}
\qvec_{n\pm1} = \qvec_n \pm h \vvec_n + \frac{h^2}{2} \, \avec_n 
\end{equation}
with $\vvec_n = M^{-1} \pvec_n$ and $\avec_n = M^{-1} \fvec_n$, we can show that 
\begin{equation}
\Phi_{n\pm1} = \Phi_n \pm h \left(\frac{\partial \Phi}{\partial \vvec}\right)_n 
+ \frac{h^2}{2} \left(\frac{\partial \Phi}{\partial \avec}\right)_n 
+ \frac{h^2}{2} \left(\vvec^T \Phi'' \vvec\right)_n + O(h^3), 
\end{equation}
where ${\partial \Phi}/{\partial \vvec}$ and ${\partial \Phi}/{\partial \avec}$ are 
given by Eq.(\ref{PHIDRV}). The simplest choice, 
\begin{equation}
\label{EXTEQ0}
\Phi^{\rm init}_{n+1} = \Phi_n = \Phi_{n+1} + O(h), 
\end{equation}
is sometimes useful, but far from satisfactory \cite{APJ}. 
A more reasonable initial guess is given by 
\begin{equation}
\label{EXTEQ1}
\Phi^{\rm init}_{n+1} = 2 \Phi_n - \Phi_{n-1} = \Phi_{n+1} + O(h^2), 
\end{equation}
which is robust and more efficient. 
Similarly, higher-order extrapolation formulae can be derived by 
using $\Phi_{n-2}, \Phi_{n-3}, ...$ \cite{APJ}. 
While these formulae give better performance at small time steps, 
instabilities occur at large time steps \cite{APJ}. 
Therefore, we usually use Eq.(\ref{EXTEQ1}) in our AIMD simulations. 

If, however, the derivatives of $\Phi$ are available, we can derive another set of formulas 
which works well for large time steps. 
For instance, when the postprocessing for atomic momenta is performed on-the-fly 
using Eq.(\ref{PPOST}), we need to calculate ${\cal H} \vvec_n$ explicitly. 
Then, we can eliminate the $O(h^2)$ term by making use of $({\partial \Phi}/{\partial \vvec})_n$ 
which is obtained as a by-product: 
\begin{equation}
\Phi^{\rm init}_{n+1} = 2 h \left(\frac{\partial \Phi}{\partial \vvec}\right)_n + \Phi_{n-1} 
= \Phi_{n+1} + O(h^3). 
\end{equation}
Alternatively, when Eqs.(\ref{RWLDS1}-\ref{RWLDS3}) are used to integrate the equations of motion, 
we need to calculate ${\cal H} \avec_n$ at each time step. 
Then, we can exploit $({\partial \Phi}/{\partial \avec})_n$ to estimate the initial guess 
with reduced $O(h^2)$ errors: 
\begin{equation}
\label{REDH2}
\Phi^{\rm init}_{n+1} = 2 \Phi_n - \Phi_{n-1} + h^2 \left(\frac{\partial \Phi}{\partial \avec}\right)_n 
= \Phi_{n+1} + O(h^2).
\end{equation}
When these advanced extrapolation schemes are used, 
the number of iterations required for electronic structure calculations 
is reduced by 10-30 \% compared to Eq.(\ref{EXTEQ1}). 
A similar approach may also be useful for geometry optimization problems \cite{MBPT3}.

\clearpage

\begin{table}[t]
\caption{
Average lengths of covalent O-H bonds obtained from molecular dynamics simulations of liquid water. 
}
\label{BONDTBL1}
\begin{center}
\begin{tabular}{ccc}
\hline
\hline
$h$ (fs) & Verlet (\AA) & Processed Verlet (\AA) \\
\hline
0.3 & 1.0300$\pm$0.0271 & 1.0300$\pm$0.0268 \\
0.6 & 1.0300$\pm$0.0273 & 1.0300$\pm$0.0270 \\
0.9 & 1.0300$\pm$0.0272 & 1.0300$\pm$0.0268 \\
1.2 & 1.0299$\pm$0.0273 & 1.0299$\pm$0.0264 \\
1.5 & 1.0300$\pm$0.0282 & 1.0299$\pm$0.0265 \\
1.8 & 1.0299$\pm$0.0290 & 1.0298$\pm$0.0264 \\
2.1 & 1.0298$\pm$0.0300 & 1.0297$\pm$0.0270 \\
\hline
\hline
\end{tabular}
\end{center}
\end{table}

\begin{table}[t]
\caption{
Average lengths of covalent O-H bonds obtained from AIMD simulations of liquid water. 
The last three lines are the results for the optimized masses.
}
\label{BONDTBL2}
\begin{center}
\begin{tabular}{ccc}
\hline
\hline
$h$ (fs) & Verlet (\AA) & Processed Verlet (\AA) \\
\hline
0.3 & 0.9854$\pm$0.0308 & - \\
0.6 & 0.9855$\pm$0.0313 & 0.9855$\pm$0.0309\\
0.9 & 0.9858$\pm$0.0321 & 0.9854$\pm$0.0307\\
1.2 & 0.9858$\pm$0.0321 & 0.9857$\pm$0.0310\\
1.0 & - & 0.9854$\pm$0.0310\\
1.5 & - & 0.9859$\pm$0.0317\\
2.0 & - & 0.9858$\pm$0.0313\\
\hline
\hline
\end{tabular}
\end{center}
\end{table}

\begin{thebibliography}{100}
\bibitem{FNOTE0} In recent years, much effort has been made to perform accurate AIMD simulations 
	using noisy forces \cite{NF1,NF2,NF3,NF4,NF5,NF6,NF7,NF8,NF9}, 
	in the spirit of Car-Parrinello fictitious dynamics \cite{NF0}. 
	At variance with these approaches, we focus on the Born-Oppenheimer molecular dynamics 
	using a small tolerance in this work.
\bibitem{FNOTE1} While Eq.(\ref{PPDEF}) is sufficient for the Verlet method, 
  the inclusion of higher-order terms is preferable 
  for higher-order integrators \cite{SIREV1,LSS1,LSS2}, as will be discussed in Sec.\ref{DISSEC}.
\bibitem{FNOTE5} We note, however, that the lack of intramolecular flexibility can 
    have a significant impact on, e.g., the phase diagram of water \cite{HAMA11}.
\bibitem{FNOTE2} To be precise, the average is taken only over the oxygen atoms 
	in the present study.
\bibitem{FNOTE3} No processing was applied in the case of $h =0.3$ fs 
    because of the high computational costs.
\bibitem{FNOTE4} In practice, $(\qvec, \pvec)$ should be replaced by $(\Qvec, \Pvec)$.
\end{thebibliography}
\end{document}